\documentclass[onecolumn,superscriptaddress,
showpacs,preprintnumbers,amsmath,amssymb]{revtex4}

\usepackage{graphicx}
\usepackage{amssymb,amsmath,amsthm}
\usepackage{bbm}

\theoremstyle{definition}

\newcommand{\bra}[1]{{\left\langle #1 \right|}}
\newcommand{\ket}[1]{{\left| #1 \right\rangle}}

\newcommand{\tr}{\mbox{$\mathrm{tr}$}}

\newcommand*{\1}{{\mathbbm{1}}}
\newcommand*{\eps}{\varepsilon}
\newcommand*{\cl}[1]{{\mathcal{#1}}}
\newcommand*{\bb}[1]{{\mathbb{#1}}}

\begin{document}

\title{Approximate quantum state sharing via two private quantum channels}

\author{Dong Pyo Chi}
\affiliation{
 Department of Mathematical Sciences,
 Seoul National University, Seoul 151-742, Korea
}
\author{Kabgyun Jeong}
\affiliation{
 Nano Systems Institute (NSI-NCRC),
 Seoul National University, Seoul 151-742, Korea
}
\date{\today}

\begin{abstract}
We investigate the approximate quantum state sharing protocol
based on random unitary channels, which is secure against
any exterior or interior attackers in principle. Although
the protocol leaks small information for a security parameter $\eps$,
the scheme still preserves its information-theoretic secrecy,
and reduces some pre-shared classical secret keys for
a private quantum channel
between a sender and two receivers.
The approximate private quantum channels constructed via
random unitary channels play a crucial role
in the proposed quantum state sharing protocol.
\end{abstract}

\pacs{
03.67.-a, 
03.67.Dd 
}
\maketitle

\section{Introduction}
Quantum physics allows us a perfect randomness, so most of all
quantum information-theoretic primitives try to offer an unconditional
security under the randomness.
For examples, quantum key distribution protocols such as
BB84~\cite{BB84} and B92~\cite{B92}
highly depend on a random measurements
for given classified non-orthogonal quantum states.

Instead of the random measurement on non-orthogonal states,
we can consider a direct randomization of quantum states
through a quantum channel.
This randomizing procedures are efficiently accomplished via
the private quantum channels~(PQC) \emph{or} quantum one-time pads~\cite{AMTW00}.
In the paper we are interest to some schemes for \emph{approximate}
encryptions~(no perfect) and we make an attempt to reducing
some classical communication resources.
We would like to call the randomizing procedures or maps as
random unitary channels~(RUC) in terms of quantum channels.
There are several methods for the approximate randomizing quantum states,
for examples,~\cite{HLSW04, AS04, DN06}: We here adapt
the procedure of Hayden et al.~\cite{HLSW04}.

Many applications of RUC in quantum protocols~(See
e.g.,~\cite{HLSW04, HHL04, BHLSW05}.)
are started from the approximate version of PQC.
Here we will propose new approximate quantum \emph{state} sharing~(AQSS) scheme,
which uses two approximate PQCs~(APQC) and reduces the
classical pre-shared secrets about one-half as compared
with a perfect protocol. Actually our protocol could be including
the (well-known) quantum \emph{secret} sharing protocols~\cite{HBB99, KKI99},
because a quantum state itself is able to operate special quantum tasks,
though those are impossible in the classical power. Imagine that
if there is a quantum computer only activated under a bipartite
quantum state~(or \emph{quantum key}), then our AQSS protocol
will give a efficient and secure solution for the quantum key.
These approximate quantum state sharing protocols may offer
us more opportunities as compared with the quantum secret sharing.

Let's take account of the pre-shared secrets for the approximate
quantum state sharing protocols under RUC-based PQC roughly.
Assume that a sender Charlie prepares a quantum state
$\varphi_{AB}$~(two-qudit) and transmits the state
through two independent RUCs, then two distant agents Alice and Bob
will receive some output state of including high entropy.
For the state $\varphi_{AB}$ the perfect
randomization protocol will require exactly the amount of
$4\log d$-unitary matrices~(Pauli matrices).
On the other hand, the construction of Hayden et al.~\cite{HLSW04}
for our AQSS scheme implies that only $2\log d+o(\log d)$-unitaries sufficient.
In other words, the perfect quantum state sharing protocol needs to
$2l$ bits of pre-shared secret information, while the AQSS protocol
demands about $l$ bits of information.
Note that the works in~\cite{AS04, DN06} will give a similar
result for $l$ bits bound.

We will prove the information-theoretic security of the AQSS scheme
in two kinds of eavesdropping: an interior and exterior attackers.
The proof of having higher entropy condition for the exterior attacks is
not easy fact, so we split the input state $\varphi_{AB}$
to separable and entangled cases. As a result,
the von Neumann entropy in both cases can be chosen sufficiently larger,
and a leakage information will be arbitrarily small.
Finally the authors show that our bipartite AQSS scheme naturally
can be generalized to an one-sender and multiparty-receivers schemes.

In section~\ref{sec:RUC}
we introduce the definition of random unitary channels,
and briefly mention about special property known as the destruction
of quantum states on a product random unitary channel.
We present our AQSS protocol based on two approximate PQCs
in section~\ref{sec:AQSS}, and investigate the security of
AQSS of considering two attacks: an exterior and interior strategies.
we finally conclude our results in section~\ref{sec:concl}.

\section{Some properties of random unitary channels} \label{sec:RUC}
Now let us define the random unitary channel, and then construct an
approximate private quantum channels.
For all density matrices $\varphi\in\cl{B}(\bb{C}^d)$, a completely positive
trace-preserving map
$\cl{N}:\cl{B}(\bb{C}^d)\to\cl{B}(\bb{C}^d)$ is the so-called
\emph{$\eps$-randomizing}, if
\begin{equation} \label{eq:epsran}
    \left\|\cl{N}(\varphi)-\frac{\1}{d}\right\|_1\le\eps,
\end{equation}
where the trace norm is defined by $\|X\|_1=\sqrt{X^\dagger X}$.
This definition directly induces the notion of random unitary channels.
That is, for every $\varphi$, a quantum channel
$\cl{N}:\cl{B}(\bb{C}^d)\to\cl{B}(\bb{C}^d)$ is called the
\emph{random unitary channel}, if
\begin{equation} \label{eq:RUC}
    \cl{N}(\varphi)=\sum_{i=1}^np_iU_i\varphi U_i^\dagger
\end{equation}
is $\eps$-randomizing, where the unitary operators $U_i\in\cl{U}(d)$,
and the probability $p_i$'s are all positives with $\sum_ip_i=1$.
(The notation $\cl{B}(\bb{C}^d)$ denotes the set of bounded linear
operators from $\bb{C}^d$ to itself and
$\cl{U}(d)\subset\cl{B}(\bb{C}^d)$ the unitary group on $\bb{C}^d$.)
Note that the parameter $n$ is the number of Kraus operation elements
for RUC, so it corresponds to the dimension of arbitrary environment.

For the approximate constructions of RUC,
it was known that for all $\eps>0$ there exist
random unitary channels
in sufficiently larger dimension $d$, such that $n$ can be taken to be
$\cl{O}(d\log d /\eps^2)$ in~\cite{HLSW04} and $\cl{O}(d/\eps^2)$ in~\cite{Aub09}
where $U_i$'s are chosen randomly according to the Haar measure.
We here fix the number $n$ of having exactly $n=\frac{150d}{\eps^2}$,
the Theorem 1 in~\cite{Aub09}.

As mentioned in the Introduction, most intuitive application of
the random unitary channel is
the \emph{approximate} private quantum channel~\cite{HLSW04},
which is a modification of the perfect private quantum
channel~\cite{AMTW00} via RUC. The RUC-based APQC is the main tool
of constructing the proposed AQSS protocol.

\begin{figure}
    \centering
    \includegraphics[scale=0.30]{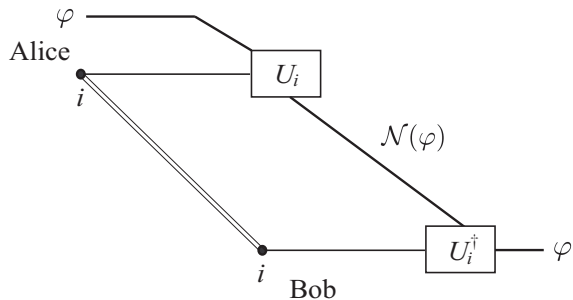}
    \caption{\label{fig:PQC}Approximate private quantum channel:
    Alice applies some $U_i$'s and Bob decodes $\cl{N}(\varphi)$
    with $i$ of having pre-shared $\log n$ bits classical information.}
\end{figure}

The security of PQC is preserved by the argument of the accessible information
in which the leakage information is less than $\eps$.
Although small information is leaked to exterior attackers,
Bob's decoding state is almost equal to Alice's original state $\varphi$.
The FIG.~\ref{fig:PQC} describes the total procedure of APQC.

In the next section we use two one-way independent PQCs
between a sender Charlie and a receiver Alice, and
the sender Charlie and another receiver Bob.
Let's define two RUCs, from the definition of ~(Eq.~(\ref{eq:RUC})), such that
\begin{eqnarray} \label{eq:2RUC}
    \cl{N}_A(\varphi):
    &=&\frac{1}{n_A}\sum_{i=1}^{n_A}U_i\varphi U_i^\dagger ~~~~\mathrm{and}\nonumber\\
    &&  \\
    \cl{N}_B(\varphi):
    &=&\frac{1}{n_B}\sum_{j=1}^{n_B}U_j\varphi U_j^\dagger, \nonumber
\end{eqnarray}
where we fix the probability as an equally weighted probabilities
$p_i=\frac{1}{n_A}$ and $p_j=\frac{1}{n_B}$ for all $i,j$, and
assume that the number of $n_A$ is equal to $n_B$,
i.e., $n_A=n_B=150d/\eps^2$.
For an approximate state sharing of any bipartite quantum state,
above two channels play an important role
in the approximate quantum state sharing scheme.

For given two RUCs $\cl{N}_A$ and $\cl{N}_B$, and for all input $\varphi_{AB}$,
we must bound the trace norm for the difference between
an output state of the product channel $\cl{N}_A\otimes\cl{N}_B$ and
maximally mixed $\1/d^2$, such that
\begin{equation} \label{eq:main}
    \left\|(\cl{N}_A\otimes\cl{N}_B)(\varphi_{AB})
    -\frac{\1_{A}\otimes\1_{B}}{d^2}\right\|_1\le\eps,
\end{equation}
where a security parameter $\eps$ be a positive less than 1.
The relation above asserts that all encoding states are
information-theoretically secure.
Unfortunately, for any entangled states proving the bound
is not a simple task.

Note that the argument
for the (efficient) randomization is related to a destruction of correlations
in quantum states~\cite{HLSW04, GPW05}. They pointed out that
the unitary operations of the amount corresponding to
the quantum mutual information
$I[A:B]=S[\varphi_A]+S[\varphi_B]-S[\varphi_{AB}]$ efficiently destroy
the total correlation of any quantum states~\cite{GPW05},
where $S[\varrho]=-\tr\varrho\log\varrho$
the von Neumann entropy. For the maximally entangled state
$\varphi_{AB}=\frac{1}{d}\sum_{i,j}\ket{ii}\bra{jj}_{AB}$,
$I[A:B]=2\log d$, which might be related to the Eq.~(\ref{eq:2logd}).

The following section gives the AQSS protocol and the security of the protocol.
The last of the section, we briefly describe a multiparty AQSS scheme.

\section{Approximate quantum state sharing protocol} \label{sec:AQSS}
Let us assume that Charlie-Alice and Charlie-Bob have independent two APQCs,
and Charlie wants to sharing a bipartite quantum state $\varphi_{AB}$
\emph{securely} between Alice and Bob.

\begin{figure}
    \centering
    \includegraphics[scale=0.30]{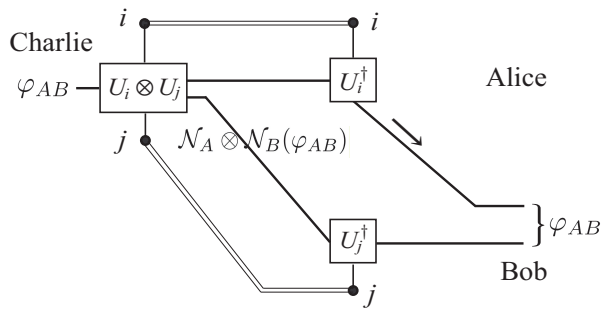}
    \caption{\label{fig:QSS3}Approximate QSS: If Charlie-Alice and Charlie-Bob
    have shared two independent PQCs each other, then the product channel
    $\cl{N}_A\otimes\cl{N}_B$ preserves the security with high probability
    for any attacks. The arrow denotes that Alice must go to Bob's location
    to obtain the state.}
\end{figure}

The protocol for a bipartite quantum state sharing is
simple~(See FIG.~\ref{fig:QSS3}):
\begin{itemize}
    \item[(i)] The sender Charlie selects a quantum state $\varphi_{AB}$
    and transmits the state through the channel $\cl{N}_A\otimes\cl{N}_B$
    to the receivers Alice and Bob.
    \item[(ii)] Distant two parties Alice and Bob just hold the state
    $\cl{N}_A\otimes\cl{N}_B(\varphi_{AB})$ they received.
    \item[(iii)] When Alice and Bob want to reveal the original
    state $\varphi_{AB}$,
    they must cooperate in a single location. They perform
    the inverse unitary operations under the locally shared keys.
\end{itemize}

The security of the AQSS protocol is divided two cases of an exterior and
interior attacks. Actually the security is based on information-theoretic
assumption, which means that the intercepted states must have the higher von Neumann
entropy. Thus any attackers cannot obtain sufficient information for
the original states.

First, let us consider an attack accomplished by an exterior Eve.
Assume that Eve intercepts the state $\cl{N}_A\otimes\cl{N}_B(\varphi_{AB})$.
We here claim that
\begin{equation} \label{eq:2logd}
    S[(\cl{N}_A\otimes\cl{N}_B)(\varphi_{AB})]\sim2\log d\to\infty
\end{equation}
as $d$ goes to infinity. We don't know the accurate description
for the state $\cl{N}_A\otimes\cl{N}_B(\varphi_{AB})$ for all inputs, so
we will divide the state $\varphi_{AB}$ into
the separable and entangled one and investigate the behavior each other.

If product state is given, it is possible to infer
the inequality Eq.~(\ref{eq:main}) easily.
By using the triangle inequality with respect to
the trace norm for the two RUCs, if
$\left\|\cl{N}_A(\varphi_A)-\frac{\1_A}{d}\right\|_1\le\eps$ and
$\left\|\cl{N}_B(\varphi_B)-\frac{\1_B}{d}\right\|_1\le\eps$,
then
$\left\|(\cl{N}_A\otimes\cl{N}_B)(\varphi_{AB})
-\frac{\1_{AB}}{d^2}\right\|_1\le2\eps$ for
$\varphi_{AB}=\varphi_A\otimes\varphi_B$.
More formally assume that
$\varphi_{AB}=\sum_ip_i\varphi_{A,i}\otimes\varphi_{B,i}$
such that $\sum_ip_i=1$,
i.e., a separable state is given, then
\begin{eqnarray}
    \left\|(\cl{N}_A\otimes\cl{N}_B)(\varphi_{AB})-\frac{\1_{AB}}{d^2}\right\|_1
    &=&\left\|\sum_ip_i\cl{N}_A(\varphi_{A,i})\otimes\cl{N}_B(\varphi_{B,i})
    -\frac{\1_{AB}}{d^2}\right\|_1 \nonumber\\
    &\le&\sum_ip_i\left\|\cl{N}_A(\varphi_{A,i})\otimes\cl{N}_B(\varphi_{B,i})
    -\frac{\1_{AB}}{d^2}\right\|_1 \label{eq:normconv}\\
    &=&\sum_ip_i\left\|\cl{N}_A(\varphi_{A,i})\otimes\cl{N}_B(\varphi_{B,i})
    -\cl{N}_A(\varphi_{A,i})\otimes\frac{\1_B}{d}+\cl{N}_A(\varphi_{A,i})\otimes\frac{\1_B}{d}
    -\frac{\1_{AB}}{d^2}\right\|_1 \nonumber\\
    &\le&\sum_ip_i\left[\left\|\cl{N}_A(\varphi_{A,i})-\frac{\1_{A}}{d^2}\right\|_1
    +\left\|\cl{N}_B(\varphi_{B,i})-\frac{\1_{B}}{d^2}\right\|_1\right] \label{eq:triangle}\\
    &\le&2\eps, \nonumber
\end{eqnarray}
where the inequalities Eq.~(\ref{eq:normconv}) and Eq.~(\ref{eq:triangle})
come from the norm convexity and the triangle inequality, respectively~\cite{HLSW04}.
Thus any separable inputs for the product channel are very close to
the maximally mixed state $\frac{\1}{d^2}$. This implies that
$S[(\cl{N}_A\otimes\cl{N}_B)(\varphi_{AB})]$ is close to $2\log d$.

For the separable input cases, there is another bound
that depends on the dimension parameter $d$ and $n$:
We can prove that the expectation value for
the difference between the channel output and the maximally mixed state
(with respect to the trace norm) is very close, that is,
\begin{equation} \label{eq:exp2ruc}
    \bb{E}_{\{U_{i,j}\}}\left\|(\cl{N}_A\otimes\cl{N}_B)(\varphi_{AB})
    -\frac{\1}{d^2}\right\|_1\le\sqrt{\frac{d^2}{n_An_B}},
\end{equation}
where $\bb{E}_{\{U_{i,j}\}}$ denotes the total expectation value of
$\{U_i\}_{i=1}^{n_A}$ and $\{U_j\}_{j=1}^{n_B}$ for the independent
RUCs $\cl{N}_A$ and $\cl{N}_B$, respectively.
The Appendix in this paper states
that the inequality Eq.~(\ref{eq:exp2ruc}) is non-trivial and obtained
precisely by exploiting the relation between the trace norm and
the Hilbert-Schmidt norm.
As mentioned above, let's take
$n_A=\frac{150d}{\eps^2}$ and $n_B=\frac{150d}{\eps^2}$, then
\begin{equation}
    \frac{d}{\sqrt{n_An_B}}=\frac{\eps^2}{150}<\eps.
\end{equation}
This implies that Eve's attack is impossible in principle.

What can we do for an entangled input state?
Though a direct proof could be impossible,
there is an evidence for the statement, the Eq.~(\ref{eq:2logd}).
The Theorem III.3 in~\cite{HLSW04} states that,
for a positive operator-valued measure~(POVM)
$\{L_i\}$ which is implemented using local operation and classical
communication~(LOCC), $\sum_i\|p_i-q_i\|_1\le\eps$, where
$p_i:=\tr(L_i(\cl{N}_A\otimes\1_B)(\varphi_{AB}))$ and
$q_i:=\tr(L_i(\frac{\1_A}{d}\otimes\varphi_B))$ with
a maximally entangled state s.t.
$\varphi_{AB}=\frac{1}{d}\sum_{i,j}^d\ket{ii}\bra{jj}_{AB}$ and
$\varphi_B=\tr_A\varphi_{AB}$.
Natural extension is possible as adding the channel $\cl{N}_B$: Define
$p_i=\tr(L_i(\cl{N}_A\otimes\cl{N}_B)(\varphi_{AB}))$ and
$q_i=\tr(L_i(\frac{\1_{AB}}{d^2}))$, then also
$\sum_i\|p_i-q_i\|_1\le\eps$. Therefore, we can conclude
the state $\cl{N}_A\otimes\cl{N}_B(\varphi_{AB})$ is close
to $\frac{\1}{d^2}$ under the LOCC-implemented POVM.
In this reason any input state $\varphi_{AB}$ through the
product channel $\cl{N}_A\otimes\cl{N}_B$ have high entropy for $d\gg1$.

Second, we must consider a situation when Alice \emph{or} Bob is malicious.
Assume that Bob intercepts the Alice's state $\cl{N}_A(\varphi_A)$,
Bob's decoded state looks like
\begin{equation}
    (\cl{N}_A\otimes\cl{N}_B^{*})(\cl{N}_A\otimes\cl{N}_B)(\varphi_{AB})
    =(\cl{N}_A\otimes\1_B)(\varphi_{AB}),
\end{equation}
where $*$ denotes the inverse operation for Bob's RUC $\cl{N}_B$,
but $S[\cl{N}_A(\varphi_A)]$ has still high entropy values.
The intercepted state $\tr_B(\cl{N}_A\otimes\1_B)(\varphi_{AB})$ is
still almost maximally mixed state by the definition of
the RUC $\cl{N}_A(\varphi_A)$.
As a result, Bob cannot obtain any information for $\varphi_A$
without Charlie-Alice's key information. Symmetrically Alice's attack is useless.
In other words, the Charlie's aim of sharing a quantum state $\varphi_{AB}$
between Alice and Bob will be securely accomplished.

At least above-mentioned two attacks~(exterior and interior eavesdropping)
cannot break the security of the proposed AQSS protocol.
so the cooperation between Alice and Bob always restores
the original state approximately.

In the proposed scenarios, the perfect protocol for quantum state sharing
requires exactly $d^4$ unitary operators, while our protocol
only needs to total $22500d^2/\eps^4$ unitaries for sufficiently larger $d$.
This fact directly means that some pre-shared key bits are reduced
by factor 2, since the AQSS is needed $2\log d-4\log\eps+\cl{O}(1)$ secret bits,
but the perfect QSS is required $4\log d$ bits.
For any state $\varphi_{AB}\in\cl{B}(\bb{C}^{d^2})$, and for any channel
$\cl{N}_{AB}$~(for an $\eps>0$ is arbitrary),
let's consider a relation like that
\begin{equation}
    \left\|\cl{N}_{AB}(\varphi_{AB})-\frac{\1}{d^2}\right\|_1\le\eps.
\end{equation}
Then, it is sufficient to construct the perfect QSS~($\eps=0$) with $d^4$
Pauli operators for the channel $\cl{N}_{AB}$ in the sense
of PQC~\cite{HLSW04, DN06}. In the case of our approximate QSS,
the product channel of
two RUCs~($\cl{N}_{AB}=\cl{N}_{A}\otimes\cl{N}_{B}$) just consume of
half secret bits, so we say that it is \emph{efficient} in weak
sense~(though small information is always leaking).

Without loss of generality,
a direct extension of the bipartite quantum state sharing
protocol~(Eq.~(\ref{eq:exp2ruc}))
gives the security of a multiparty approximate quantum state sharing~(MAQSS).
Assume that a sender Charlie~($C$) prepares
an $m$-qudit $\varphi_{A_1A_2\cdots A_m}$.
If they initially have shared PQCs between $C$-$A_1$, $C$-$A_2$ and so on,
then, for any $\eps>0$,
\begin{equation} \label{eq:MAQSS}
    \left\|(\cl{N}_{A_1}\otimes\cdots\otimes\cl{N}_{A_m})(\varphi_{A_1A_2\cdots A_m})
    -\frac{\1}{d^{m}}\right\|_1\le\eps.
\end{equation}
The above Eq.~(\ref{eq:MAQSS}) implies that any exterior attacks will be failed.
Furthermore all interior attacks~(including group conspiracy) will be frustrated
to obtain the whole state without others secrets,
it has similar reason to the two receivers protocol.
Let's look at the cost of secret bits for the MAQSS scheme.
Roughly speaking, the perfect scheme requires $2m\log d$ secret bits,
but MAQSS only $m\log d+o(\log d)$-bits sufficient.

\section{Conclusions} \label{sec:concl}
We studied that the approximate quantum state sharing schemes
are efficient from the classical information cost of view and
those are robust to the two kinds of attacks.
The proposed AQSS protocol basically depends on
an approximate private quantum channel,
which is constructed via two independent random unitary channels.
Although the protocol leaks small information corresponding to
the security parameter $\eps$,
the scheme preserves its information-theoretic security, and
so the AQSS and MAQSS schemes can be interpreted as
some high-efficiency state sharing protocols
for any bipartite and multipartite quantum states.

\acknowledgements{
This work was supported by Basic Science Research Program
through the National Research Foundation of Korea~(NRF) funded
by the Ministry of Education, Science and Technology~(Grant No. 2009-0072627).}

\section*{Appendix}
For given two random unitary channels
$\cl{N}_A(\varphi)$ and
$\cl{N}_B(\varphi)$ in Eq.~(\ref{eq:2RUC}),
and for all pure-separable states
$\varphi_{AB}\in\cl{B}(\bb{C}^{d^2})$,
\begin{eqnarray}
    \left\|(\cl{N}_A\otimes\cl{N}_B)(\varphi_{AB})\right\|_2^2
    &=&\tr(\cl{N}_A\otimes\cl{N}_B)^2(\varphi_{AB}) \nonumber\\
    &=&\frac{1}{n_A^2n_B^2}\sum_{i=1}^{n_A}\sum_{j=1}^{n_B}
    \tr\left(U_i\otimes U_j\varphi_{AB}U_i^\dagger\otimes U_j^\dagger\right)^2 \nonumber\\
    &&+\frac{1}{n_A^2n_B^2}\sum_{i\neq k}^{n_A}\sum_{j\neq l}^{n_B}\tr
    \left(U_i\otimes U_j\varphi_{AB}U_i^\dagger\otimes U_j^\dagger\right)
    \left(U_k\otimes U_l\varphi_{AB}U_k^\dagger\otimes U_l^\dagger\right),
\end{eqnarray}
where $\tr\left(U_i\otimes U_j\varphi_{AB}U_i^\dagger\otimes U_j^\dagger\right)^2=1$
for any pure states $\varphi_{AB}$.
(Note that this method is just an extension of the statement, the chapter 3
in~\cite{DN06}.)

Recall that the unitary operators are chosen randomly according to the Haar measure,
and take expectation over the random selection of unitaries:
\begin{eqnarray}
    \bb{E}_{\{U_{i,j}\}}\left[\tr(\cl{N}_A\otimes\cl{N}_B)^2(\varphi_{AB})\right]
    &=& \frac{1}{n_An_B} \nonumber\\
    &&+\frac{1}{n_A^2n_B^2}\sum_{i\neq k}\sum_{j\neq l}
    \bb{E}_{\{U_{i,j}\}}\tr
    \left(U_i\otimes U_j\varphi_{AB}U_i^\dagger\otimes U_j^\dagger\right)
    \left(U_k\otimes U_l\varphi_{AB}U_k^\dagger\otimes U_l^\dagger\right) \nonumber\\
    &=& \frac{1}{n_An_B}+\tr\left[\bb{E}_{\{U_{i,j}\}}
    \left(U_i\otimes U_j\varphi_{AB}U_i^\dagger\otimes U_j^\dagger\right)
    \bb{E}_{\{U_{k,l}\}}
    \left(U_k\otimes U_l\varphi_{AB}U_k^\dagger\otimes U_l^\dagger\right)\right]
    \label{eq:exp1}\\
    &=&\frac{1}{n_An_B}+\tr\frac{\1}{d^4} \label{eq:exp2}\\
    &=&\frac{1}{n_An_B}+\frac{1}{d^2}. \label{eq:exp3}
\end{eqnarray}
In Eq.~(\ref{eq:exp1}), we have used that $U_{i,j}$ and $U_{k,l}$ are chosen
independently, and Eq.~(\ref{eq:exp2}) inherited from the definition
of the Haar measure. (For any $\varphi\in\cl{B}(\bb{C}^d)$, a Haar-distributed set
$U:=\{U_i\}_{i=1}^n$ satisfies that $\bb{E}_UU\varphi U^\dagger
=\int U\varphi U^\dagger dU=\frac{\1}{d}$.) The Eq.~(\ref{eq:exp2})
exploits the separable condition for $\varphi_{AB}$.
Note that for any rank $d$ matrix $X$ $\|X\|_1\le\sqrt{d}\|X\|_2$.
For any rank $d^2$ matrix $X$,
a generalization of the Corollary A.2 in~\cite{DN06} directly show that
\begin{equation}
    \left\|X-\frac{\1_A\otimes\1_B}{d^2}\right\|_1^2
    \le d^2\|X\|_2^2-1.
\end{equation}

Then, from considering the random variable
$Y:=\left\|(\cl{N}_A\otimes\cl{N}_B)(\varphi_{AB})-\frac{\1}{d^2}\right\|_1$
and Eq.~(\ref{eq:exp3}),
\begin{eqnarray}
    \bb{E}Y
    &\le& \sqrt{\bb{E}Y^2} \nonumber\\
    &\le& \sqrt{d^2\|Y\|_2^2-1} \\
    &=& \sqrt{\frac{d^2}{n_An_B}}.\nonumber
\end{eqnarray}


\end{document}